\documentclass[twocolumn,11pt]{article}

\usepackage[letterpaper,margin=0.5in,top=1in,bottom=1in]{geometry}
\usepackage{amsfonts,epsfig,amsmath,amsthm,amssymb,graphics,verbatim,overpic}

\def\R{\mathbb{R}}

\def\N{\mathbb{N}}

\def\P{\mathbb{P}}

\def\cC{\mathcal{C}}

\def\cI{\mathcal{I}}

\def\cN{\mathcal{N}}
\def\cO{\mathcal{O}}

\def\ra{\rightarrow}
\def\I{\infty}

\newcommand{\be}{\begin{equation}}
\newcommand{\ee}{\end{equation}}
\newcommand{\benn}{\begin{equation*}}
\newcommand{\eenn}{\end{equation*}}
\newcommand{\bea}{\begin{eqnarray}}
\newcommand{\eea}{\end{eqnarray}}
\newcommand{\beann}{\begin{eqnarray*}}
\newcommand{\eeann}{\end{eqnarray*}}

\def\txta{{\textnormal{a}}}

\def\txtd{{\textnormal{d}}}
\def\txte{{\textnormal{e}}}

\def\txtu{{\textnormal{u}}}
\def\txtr{{\textnormal{r}}}

\def\ra{\rightarrow}
\def\I{\infty}

\title{Quenched Noise and Nonlinear Oscillations\\ in Bistable Multiscale Systems}

\author{Christian Kuehn\\ Technical University of Munich (TUM)}

\begin{document}

\maketitle

\textbf{Abstract:}  
{\it Nonlinear oscillators are a key modelling tool in many applications.
The influence of annealed noise on nonlinear oscillators has been studied 
intensively. It can induce effects in nonlinear oscillators not present in the deterministic 
setting. Yet, there is no theory regarding the quenched noise scenario of random 
parameters sampled on fixed time intervals, although this situation is often a lot
more natural. Here we study a paradigmatic nonlinear oscillator of 
van-der-Pol/FitzHugh-Nagumo type under quenched noise as a piecewise-deterministic
Markov process. There are several interesting effects such as period
shifts and new different trapped types of small-amplitude oscillations, which can be
captured analytically. Furthermore, we numerically discover
quenched resonance and show that it differs significantly from previous finite-noise 
optimality resonance effects. This demonstrates that quenched oscillators
can be viewed as a new building block of nonlinear dynamics.}\medskip

Classical oscillators are of fundamental importance for many phenomena, either
on an individual level~\cite{GH,Nayfeh}, or in a coupled-oscillator case 
for complex networked dynamics systems~\cite{Strogatz2}. The most frequently 
encountered nonlinear oscillator are van der Pol~\cite{vanderPol1} (or 
FitzHugh-Nagumo~\cite{FitzHugh}) bistable systems, which have found applications 
in mechanical systems, neuroscience, lasers, electronics, chemistry, systems biology, 
geophysics, and many other areas, where multiscale hysteresis oscillation patterns 
are observed. Beyond classical hysteresis-type or relaxation 
oscillations~\cite{vanderPol1},
there are more complex oscillation patterns in forced oscillations such as mixed-mode 
oscillations (MMOs)~\cite{Desrochesetal} composed of small-amplitude oscillations (SAOs)
and large-amplitude oscillations (LAOs). Furthermore, the influence of noise on
oscillations has been a significant development, including the description
of effects such as stochastic resonance (SR)~\cite{BenziSuteraVulpiani,NicolisNicolis}, 
coherence resonance (CR)~\cite{PikovskyKurths}, self-induced stochastic 
resonance~\cite{MuratovVanden-EijndenE}, steady-state stochastic 
resonance~\cite{KuehnNetworks} and inverse stochastic resonance~\cite{GutkinJostTuckwell}; 
see~\cite{GammaitoniHaenggiJungMarchesoni}
for a broad review of SR. The underlying, and unexpected, principle of these noise-induced
resonances is that a \emph{finite} noise strength optimizes the oscillation
regularity. Yet, these effects have been developed and studied under the assumption of
\emph{annealed disorder}, leading to stochastic forces such as colored, white, or 
shot-noise~\cite{Gardiner}. There are also studies of annealed noise MMOs, 
e.g.~\cite{BerglundLandon,SimpsonKuske}. 
Here we consider the \emph{quenched disorder} case for multiscale nonlinear oscillators.
In fact, several stochastic effects are modelled by random 
parameters, which just change at certain time instances, rather than parameters fluctuating 
continuously. A simple example are mechanical systems involving random loading. A load usually
does not change continuously but at certain time instances. Quenched disorder naturally leads 
to a piecewise-deterministic Markov process~\cite{Davis}. It also opens 
up new connections to uncertainty quantification~\cite{Grigoriu2} adding a fundamental 
nonlinear component for uncertainty propagation. We also remark that we study
\emph{quenched noise} in a single oscillator. Hence, this work is not directly connected 
to quenched \emph{oscillation and amplitude death}~\cite{KosekaVolkovKurths}, which 
are phenomena present already in (coupled) deterministic oscillator models; yet, it 
would be highly interesting
if it is possible to establish potential connections in the future.\medskip

Our main results are three 
phenomena: period modulation for 
LAOs, two new types of SAOs, and the main novel effect of quenched resonance over a broad 
range of noises including start and termination effects. We provide analytical as well
as numerical evidence for our results and show that the choice of quenched noise 
distribution can be very broad, illustrating the effects for the uniform, exponential
and normal distributions; the results are easily adapted to far broader classes of 
distributions. In summary, this Letter demonstrates the broad potential for 
quenched oscillators as a new building block of nonlinear dynamics.\medskip

\textbf{Setting:} The prototypical example we are going to study is the 
quenched-random forced van der Pol oscillator
\be
\label{eq:vdP2nd}
\frac{\txtd^2 x}{\txtd t^2} + \mu(x^2-1)\frac{\txtd x}{\txtd t}+x=
P(t,\omega), 
\ee
where $x=x(t)\in\R$, $\mu>0$ controls the (nonlinear) damping, we 
assume $\mu \gg 1$, and $P=P(t,\omega)$ is a quenched-random, time-dependent 
forcing, i.e., $P$ is going to be fixed for time intervals $\cI_j=[t_{j+1},
t_j)\subset [0,\I)$ with $j\in\N$, and $P$ switches randomly at the 
beginning of each time interval to a new value according to a given 
distribution. We set $(\Delta t)_j:=t_{j+1}-t_j$ and 
assume true quenching, i.e., $t=\cO(1)$, thereby avoiding 
any rapidly-switched forcing limits. The random forcing 
we consider can easily be implemented experimentally in mechanical 
systems. It also occurs naturally in neuroscience, where neuron
model parameters are fixed for considerable time scales, yet may 
change due to internal or external events. The second-order 
equation~\eqref{eq:vdP2nd} can be transformed via a 
standard Li\'enard ansatz 
\benn
y:=\frac{1}{\mu^2}\frac{\txtd x}{\txtd t}+\frac13 x^3-x, \quad
\varepsilon:=1/\mu^2,\quad  s:=t/\varepsilon,
\eenn 
into a fast-slow system ($0<\varepsilon\ll 1$)
\be
\label{eq:vdP}
\begin{array}{rcrcl}
\varepsilon \frac{\txtd x}{\txtd s} &=& \varepsilon \dot{x} &=& 
y-\frac13 x^3+x=:f(x,y),\\
\frac{\txtd y}{\txtd s} &=& \dot{y} &=& P-x=:g(x,y).
\end{array}
\ee
Since $P=P(t,\omega)=P(\varepsilon s,\omega)$ and time intervals with 
$(\Delta t)_j=\cO(1)$ yield $(\Delta s)_j=\cO(1/\varepsilon)$, there are long
epochs of order $\cO(1/\varepsilon)$, where the system is completely 
deterministic on the slow time scale $s$. Of course, one may also 
re-write~\eqref{eq:vdP} on the fast time scale $t=s/\varepsilon$
\be
\label{eq:vdP1}
\begin{array}{rcrcl}
\frac{\txtd x}{\txtd t} &=& x' &=& y-\frac13 x^3+x,\\
\frac{\txtd y}{\txtd t} &=& y' &=& \varepsilon (P-x).
\end{array}
\ee
We recall the fast-slow bifurcation analysis of the
van der Pol oscillator to fix the notation~\cite{KuehnBook}. The critical manifold is 
$\cC_0:=\{(x,y)\in\R^2:y=\frac13 x^3-x=:h(x)\}$. It consists of steady
states for the fast subsystem
\be
\label{eq:vdP2}
\begin{array}{rcl}
x' &=& y-\frac13 x^3+x,\\
y' &=& 0,
\end{array}
\ee
obtained from~\eqref{eq:vdP1} in the singular limit $\varepsilon\ra 0$. 
$\cC_0$ contains two fold points $p_\pm=(\pm 1,\mp 2/3)$, which separate
the manifold into attracting parts $\cC^{\txta-}_0=\cC_0\cap \{x<-1\}$, 
$\cC^{\txta-}_0=\cC_0\cap \{x>1\}$, and the repelling part 
$\cC^{\txtr}_0=\cC_0\cap \{-1<x<1\}$. The three parts are normally 
hyperbolic~\cite{Fenichel4} and have the claimed stability properties 
for~\eqref{eq:vdP2} since $\partial_x f(p)< 0$ for $p\in\cC_0^{\txta\pm}$ 
and $\partial_x f(p)> 0$ for $p\in\cC_0^{\txta\pm}$. $\cC_0$ can also be 
viewed as the phase space of the slow subsystem 
\be
\label{eq:vdP3}
\begin{array}{rcl}
0 &=& y-\frac13 x^3+x,\\
\dot{y} &=& P-x,
\end{array}
\ee
which is a differential-algebraic equation (DAE) arising as the singular 
limit of~\eqref{eq:vdP}. If $P\in\R$ is viewed as 
a \emph{fixed deterministic parameter}, there are two fold bifurcations 
in the fast subsystem at $P=\pm 1$, which correspond to singular Hopf
bifurcation points of the full system~\eqref{eq:vdP}~\cite{KuehnBook}. 
In this context, the unique global steady state $(x^*,y^*)=(P,h(P))$ is 
globally stable for $|P|>1$ and destabilizes at two singular Hopf 
bifurcations leading to large-amplitude relaxation oscillations via 
a canard explosion~\cite{Diener}. If $P(\cdot,\omega)$ is chosen randomly 
at multiple times $t_j$, then physical intuition and results from SR/CR
suggest that oscillations can be \emph{mixtures} of \emph{near-stationary} and
\emph{oscillatory}. However, due to the quenching, the dynamical 
phase space geometry and statistical properties of these mixtures 
are very different from established phenomena. We are going to analyze
quenched nonlinear oscillations in three steps: (I) random modulations 
of the period of relaxation oscillations (or pure LAOs), (II) generation 
of different types of SAOs, and (III) the analysis of resonance phenomena 
within MMOs.\medskip  

\begin{figure}[htbp]
	\centering
\begin{overpic}[width=1\linewidth]{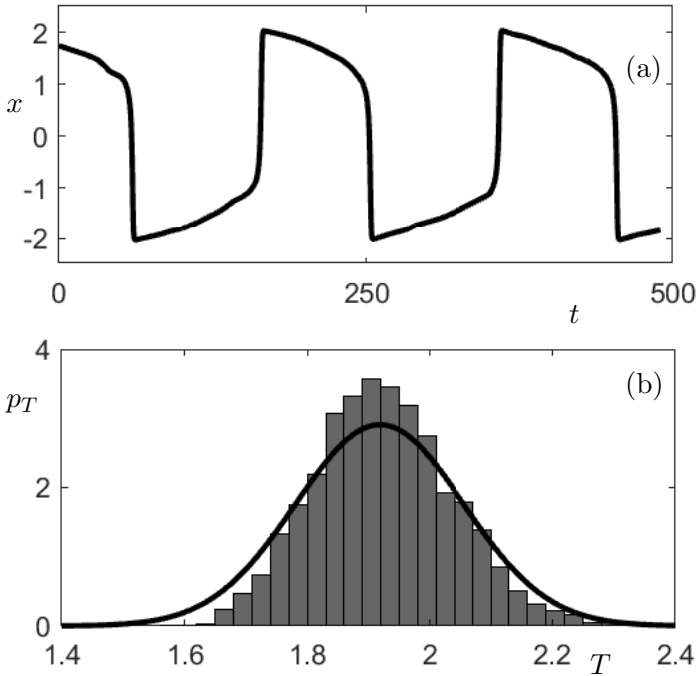}
\put(88,85){(a)}
\put(88,40){(b)}
\put(80,50){$t$}
\put(0,80){$x$}
\put(83,0){$T$}
\put(0,38){$p_T$}
\end{overpic}
\caption{\label{fig:01}(a) $x$-time series of~\eqref{eq:vdP} drawn up to $T=500$; 
full trajectory is of length $T=5\cdot 10^{6}$. Parameters are $\varepsilon=0.01$, 
$l=0.75$ and $(\Delta t)=10$. (b) Modulated period histogram computed from the 
full trajectory shifted by $3.5\cdot \varepsilon^{2/3}$; solid curve is a Gaussian
density calculated from the analytical random Dorodnityn-type approximation.}
\end{figure}

\textbf{(I) Quenched LAOs:} We fix the duration between switches uniformly
$(\Delta t)_j = (\Delta t)$ for all $j$; other cases are possible but
will not be discussed here. We assume $P$ at $t_j$ 
is sampled from a density $p=p(\omega)$, $\omega\in\R$. From standard 
fast-slow systems results~\cite{KuehnBook}, it follows that if 
the support of $p$ is compact and properly contained in the 
interval $(-1,1)$, then there exists $\varepsilon_0>0$ such that 
for all $\varepsilon\in(0,\varepsilon_0]$ we get standard relaxation 
oscillations; see also Figure~\ref{fig:01}(a). Yet, the \emph{period} 
$T$ will be a random variable $T=T(\omega)$ depending upon the chosen 
distribution $p=p(\omega)$. Since \eqref{eq:vdP} remains 
invariant under the symmetry $x\mapsto -x$, $y\mapsto -y$, relaxation
oscillations are symmetric so we assume for now $p(\omega)=p(-\omega)$. 
To leading-order $T=T_0+\cO(\varepsilon^{2/3})$~\cite{MisRoz}, where $T_0$ 
is approximated by the slow drift on $\cC_0^{\txta\pm}$ via the slow 
subsystem~\eqref{eq:vdP3}. Implicit time differentiation gives
\benn
\dot{y}=\dot{x}(x^2-1)\quad\Rightarrow \dot{x} = \frac{P-x}{x^2-1}.
\eenn
A Dorodnitsyn-type~\cite{MisRoz,KuehnBook} calculation yields
\benn
T=T_0+\cO(\varepsilon^{2/3})\approx
\int_{2}^1\frac{x^2-1}{P-x}~\txtd x+\int_{-2}^{-1}\frac{x^2-1}{P-x}~\txtd x,
\eenn
where we used a leading-order drop point approximation for the random 
starting point on $\cC_0^{\txta\pm}$ by just setting $x_0(\omega)\approx 
\pm 2$ and dropped the perturbation term $\cO(\varepsilon^{2/3})$. Working 
out the integral yields
\beann
T_0&=&
2 P^2 \coth ^{-1}(2 P+3)+2 P^2 \coth ^{-1}(3-2 P)+\\
&&\ln 
\left(\frac{P-1}{P-2}\right)-\ln \left(\frac{P+2}{P+1}\right)+3=:v(P).
\eeann

Since $P\in(-1,1)$ throughout this section, the function $v:(-1,1)\ra 
(v(-1),v(+1))=:(q_-,q_+)$ is symmetric, i.e.~$v(P)=v(-P)$. Furthermore,
$v$ is smooth and piecewise monotonic with $v'<0$ for $P\in(-1,0)$ and 
$v'>0$ for $P\in(0,1)$. So $P$ having density $p$, 
gives that $T$ has approximately density $q(y)=p(v^{-1}(y))(v^{-1})'(y)$,
where the inverse is understood in a piecewise sense. 
Calculating $v^{-1}$ explicitly is not possible, yet calculating centered 
moments is easy since
\beann
\langle (T-\mu)^k\rangle &=&\int_{q_-}^{q_+} (y-\mu)^k~ p(v^{-1}(y)) ~v^{-1})'(y)~\txtd y\\ 
&=& \int_{-1}^{1}(v(\omega)-\mu)^k~p(\omega)~\txtd \omega,\quad \mu:=\langle T\rangle.
\eeann  
Even for the basic distributions $p$, we get interesting results.
Let $p_\delta(x)=\delta(x)$ be a point mass, let $p_{\txtu,l}$ be a uniform 
distribution on $(-l,l)$ with $l\in(0,1)$. Then $\mu_\delta=3-2\ln 2$ is
the classical deterministic result. For the uniform case, the integrals 
can be evaluated explicitly, yet the formula is lengthy. The second-order
Taylor polynomial near $l=0$ of the explicit formula gives 
\benn
p_{\txtu,l}=(3-2\ln (2))+l^2 \left(\frac{\log (4)}{3}-\frac{1}{4}\right)+\cO(l^4). 
\eenn
so the average period increases as the distribution broadens, which actually
is an observation true irrespective of the Taylor approximation. One may easily 
evaluate the variance $\sigma^2_\txtu$, and higher-order moments, 
from the formulas above. For very long trajectories or relatively small $(\Delta t)$,
we expect by Central Limit Theorem to obtain a normal distribution 
$T_0\ra \mathcal{N}(\mu_{\txtu,l},\sigma_{\txtu,l})$; see~Figure~\ref{fig:01}(b). 
We observe that the largest error in the calculation arises by neglecting 
the term $\cO(\varepsilon^{2/3})$. Short/intermediate-term fluctuations are 
non-Gaussian for generic input distributions. In general, input noise 
in $P$ gets \emph{transformed} to noise in the period via a \emph{nonlinear} 
mapping $v$.\medskip
 
\begin{figure}[htbp]
	\centering
\begin{overpic}[width=1\linewidth]{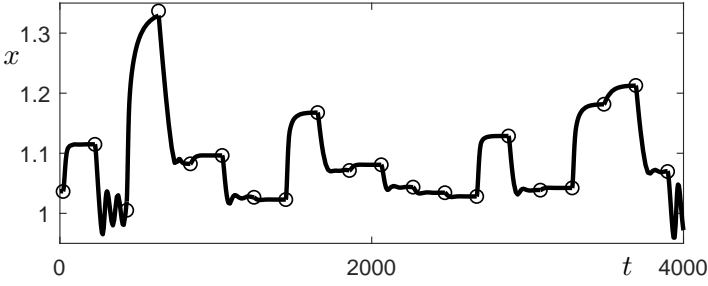}
	\put(88,0){$t$}
	\put(0,30){$x$}
\end{overpic}
\caption{\label{fig:02}$x$-time series of~\eqref{eq:vdP} with $\varepsilon=0.01$, 
and $P$ drawn from the shifted exponential distribution with $\mu=0.1$ and 
$(\Delta t)=200$.}
\end{figure}

\textbf{(II) Quenched SAOs:} 
Relaxation oscillations can also be completely blocked, despite quenched noise,
in the case $P\in(-\I,-1)$ or $P\in (1,\I)$. We assume here $P\in(1,\I)$, consider 
an initial condition on $\cC^{\txta+}_0$ but $\varepsilon>0$ and now take a shifted 
exponential density $p(\omega)=\frac1\mu \txte^{-(\omega-1)/\mu}$ for $\omega>1$,  
and $p(\omega )=0$ zero otherwise. A time series is shown in Figure~\ref{fig:02}.

We observe that there are \emph{two} types of SAOs.
On the one hand, we get small oscillations due to a stable spiral steady state 
for~\eqref{eq:vdP}. The condition can be calculated by linearizing at 
$(x_*,y_*)=(P,h(P))$, considering the Jacobian $J$, and using the trace-determinant
condition for complex eigenvalues
\benn
\textnormal{trace}(J)^2-4\det(J)<0\quad \Leftrightarrow \quad (1-P^2)^2
<4\varepsilon.
\eenn
For our choice of exponential distribution, or similarly for other distributions, we
now just have to calculate under the assumption that $(\Delta t)$ is sufficiently large
to track the equilibrium that
\beann
&\mathbb{P}(\textnormal{spiral SAO in $(t_j,t_{j+1})$})=\int_{1}^{\sqrt{1+\sqrt{4\varepsilon}}} 
\frac1\mu \txte^{-(\omega-1)/\mu}~\txtd \omega\\
&=
1-\txte^{\frac{1-\sqrt{2 \sqrt{\epsilon }+1}}{\mu }}
=\frac{\sqrt{\epsilon }}{\mu }+\frac{(-\mu -1) \epsilon }{2 \mu ^2}+\cO(\varepsilon^{3/2}).
\eeann
Therefore, there is a very fine balance between the time scale separation
and the mean of the exponential distribution for small spiral SAOs. 

\begin{figure}[htbp]
	\centering
\begin{overpic}[width=1\linewidth]{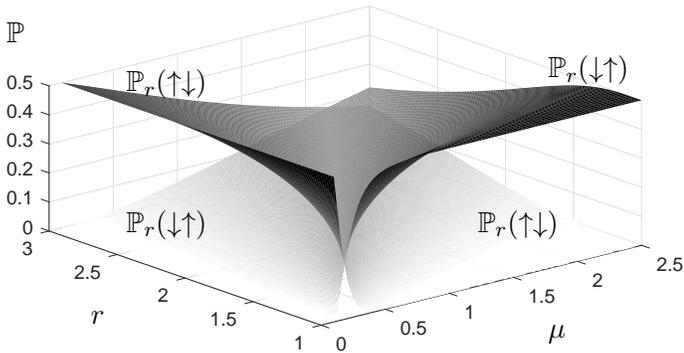}
	\put(80,3){$\mu$}
	\put(15,5){$r$}
	\put(3,45){$\P$}
	\put(20,18){$\P_r(\downarrow\uparrow)$}
	\put(80,40){$\P_r(\downarrow\uparrow)$}
	\put(20,38){$\P_r(\uparrow\downarrow)$}
	\put(70,18){$\P_r(\uparrow\downarrow)$}
\end{overpic}
\caption{\label{fig:05}Probability of non-spiral half-rotation SAOs showing the two 
intersecting smooth surfaces $\P_r(\downarrow\uparrow)$ and $\P_r(\uparrow\downarrow)$.}
\end{figure}

If we just focus on SAOs without small spirals, then a good
approximation is just to let $\varepsilon=0$ and calculate the shifting pattern
of the random steady state $x^*_j=P(t_j,\omega)\in\cC_0^{\txta+}$ for the slow
subsystem~\eqref{eq:vdP3}. Let 
\benn
\P_r(\downarrow\uparrow):=\P(x^*_{j+1}<x^*_j,x^*_{j+1}<x^*_{j+2}|x^*_j=r),
\eenn
i.e., it is the probability that we start at level $r$, then move down slowly very 
close to the new steady state (assuming $(\Delta t)$ is sufficiently large), and then
moving up towards the next steady state. Hence, $\P_r(\downarrow\uparrow)$ is the 
probability to get a non-spiral half-oscillation SAO within $2(\Delta t)$. Similarly, 
we can define and interpret $\P_r(\uparrow\downarrow)$. We calculate for the exponential
distribution case, setting $p_{1r}:=\int_1^rp(\omega)~\txtd \omega$, that
\beann
\P_r(\downarrow\uparrow)&=& \int_1^r p_{1r}(1-p_{1s})p(s)~\txtd s \\
&=& \frac{1}{2} \txte^{-\frac{3 r}{\mu }} \left(\txte^{1/\mu }-
\txte^{r/\mu }\right)^2 \left(\txte^{1/\mu }+\txte^{r/\mu }\right).
\eeann
Similarly, we calculate 
\beann
\P_r(\uparrow\downarrow)&=& \int_r^\I (1-p_{1r})p_{1s} p(s)~\txtd s \\
&=& \txte^{\frac{2-2 r}{\mu }}-\frac{1}{2} \txte^{\frac{3-3 r}{\mu }}.
\eeann
Figure~\ref{fig:05} shows the results as functions of $r$ and 
$\mu$. Similarly, one can also deal with more complex SAO sequences, and
calculate the probabilities $\P_r(\uparrow\uparrow\downarrow), 
\P_r(\downarrow\downarrow\uparrow)$, etc of patterns explicitly. If
$(\Delta t)$ is too small, we do not track the steady states and even 
more complex patterns arise, which are captured by the precise time 
evolution of the slow subsystem similar to situation for LAOs discussed 
above.\medskip   

\textbf{(III) Quenched Resonance:} 
Instead of modulated relaxation oscillations, which can also be viewed as 
LAOs, and SAOs as discussed in the previous
section, we can also expect \emph{mixtures} of these under the influence
of noise. In this section, we are sampling from a normal distribution 
$P\sim \cN(\mu,\sigma^2)$.

\begin{figure}[htbp]
	\centering
\begin{overpic}[width=1\linewidth]{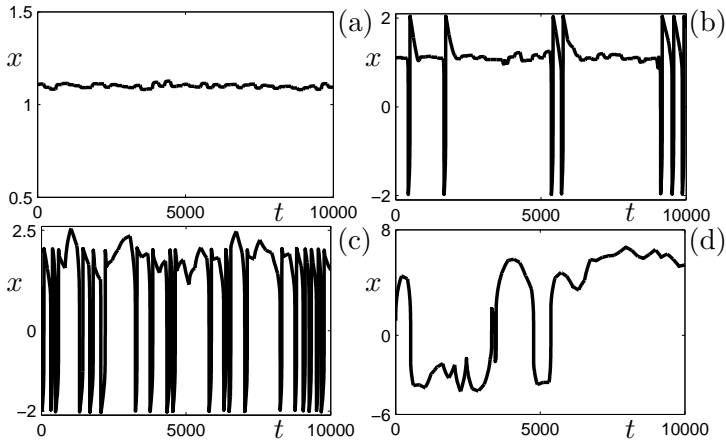}
	\put(88,0){$t$}
	\put(38,0){$t$}
	\put(0,21){$x$}
	\put(0,53){$x$}
	\put(51,21){$x$}
	\put(51,53){$x$}
	\put(38,31){$t$}
	\put(88,31){$t$}
	\put(47,58){(a)}
	\put(97,58){(b)}
	\put(47,27){(c)}
	\put(97,27){(d)}
\end{overpic}
\caption{\label{fig:03}Time series with $\varepsilon=0.01$, and $P$ drawn from a 
Gaussian distribution with $\mu=0.1$, and we fixed $(\Delta t)=200$. 
(a) $\sigma=0.01$. (b) $\sigma=0.1$. (c) $\sigma=1$. (d) $\sigma=10$.}
\end{figure}

Figure~\ref{fig:03} shows examples of the possible dynamics. 
The mean steady state is located at $(x^*,y^*)=(\mu,h(\mu))=(1.1,h(1.1))$ for 
our choice of parameters. Four different regimes can be observed upon increasing 
the noise level: (A) no LAOs/spikes but some SAOs as shown in 
Figure~\ref{fig:03}(a); (B) occasional LAOs interspersed with SAOs, i.e., MMOs 
occur as shown in Figure~\ref{fig:03}(b); (C)
MMOs with more regular LAOs are displayed in Figure~\ref{fig:03}(c); (D) some 
large spikes but also long periods of random  drift near the attracting 
branches $\cC_0^{\txta\pm}$ are shown in Figure~\ref{fig:03}(d). To quantify
the regularity of the oscillations, we use the noise-to-signal ratio known
from SR/CR quantification \cite{PikovskyKurths,LindnerSchimansky-Geier}
\benn
R=\frac{\sqrt{\langle \tau^2\rangle-\langle \tau\rangle ^2}}{\langle \tau\rangle},
\eenn 
where $\tau$ is the duration between two zero crossings of $x$ with $x'>0$ at 
the crossing.

\begin{figure}[htbp]
	\centering
\begin{overpic}[width=1\linewidth]{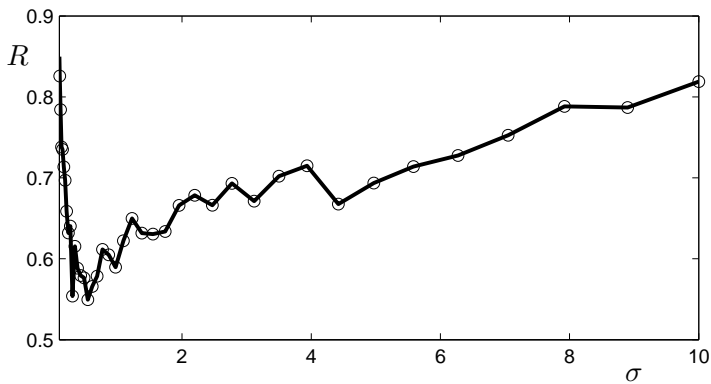}
	\put(88,0){$\sigma$}
	\put(0,45){$R$}
\end{overpic}
\caption{\label{fig:04}Standard deviation $\sigma$ of the input forcing 
distribution versus noise-to-signal ratio. Parameters are $\varepsilon=0.01$, 
$\mu=0.1$, and $(\Delta t)=200$. For the sampling we considered $20$ sample 
paths of slow-time length $s=1000$ on a logarithmically-spaced $\sigma$-space
(indicated by circles), and interpolated the measurement points linearly.}
\end{figure}

Our intuition from Figure~\ref{fig:03} is confirmed with detailed
statistics of $\sigma$ and $R$ in Figure~\ref{fig:04}. For small noise level
$\sigma$, there are only rare excursions as the system mostly has a steady
state on $\cC_0^{\txta+}$. For a very broad range of intermediate noises, the
behaviour improves as $R$ decreases implying more regular LAOs, while for very
large noise level, there is again a higher probability of getting trapped on
either of the two branches $\cC_0^{\txta\pm}$ for extended time periods.
In particular, there is a finite \emph{optimal noise level} for the input
forcing, which we refer to as \emph{quenched resonance}. There is some similarity 
to CR/SR/etc as one has to eliminate via noise 
a deterministically stable steady state. Yet there are substantial differences 
to CR/SR/etc. For low noise levels, there are \emph{no} oscillations. Furthermore,
there is a \emph{much broader range of noises}, where an improved regular 
oscillation can be observed. Lastly, the termination mechanism does not
lead to a purely noisy but to a two-stage \emph{nearly trapped} 
piecewise-deterministic process. \medskip

\textbf{Summary:} We have shown that quenched disorder leads to new effects 
in the core class of multiscale nonlinear oscillators. This applies to slow 
scales for SAOs and to fast-slow scales for LAOs. We have proven that analytical 
calculations are possible and can be carried out for different random 
forcings. Furthermore, quenched resonance was discovered, which differs 
significantly from CR/SR. However, this Letter is clearly just a starting point 
for a far more general study of nonlinear multiscale dynamical systems under 
quenched disorder, on the individual as well as network level.\medskip

\textbf{Acknowledgements:} 
I would like to thank the VolkswagenStiftung for support via a 
Lichtenberg Professorship.

{
\bibliographystyle{unsrt}
\bibliography{../my_refs}
}

\end{document}